\documentclass[pra,aps,onecolumn,nofootinbib,superscriptaddress]{revtex4}
\usepackage[T1]{fontenc}

\usepackage{amsmath,amssymb}
\usepackage{amsfonts}
\usepackage{epsfig,pstricks,graphicx}
\usepackage{bm}
\usepackage{physics}

\definecolor{armygreen}{rgb}{0.29, 0.33, 0.13}
\usepackage{color}
\usepackage{array}
\def\id{{\rm 1\kern-.22em l}}

\newcommand{\mm}[1]{{\color{black}  #1}}
\usepackage[normalem]{ulem} 
\usepackage{soul} 
\usepackage{epigraph}

\begin{document}

\title{Comment on `Single particle nonlocality with completely
independent reference states'}
 
\author{Tamoghna Das}

\affiliation{International Centre for Theory of Quantum Technologies, University of Gdańsk, Wita Stwosza 63, 80-308 Gdańsk, Poland}

\author{Marcin Karczewski}

\affiliation{International Centre for Theory of Quantum Technologies, University of Gdańsk, Wita Stwosza 63, 80-308 Gdańsk, Poland}

\author{Antonio Mandarino}

\affiliation{International Centre for Theory of Quantum Technologies, University of Gdańsk, Wita Stwosza 63, 80-308 Gdańsk, Poland}

\author{Marcin Markiewicz}

\affiliation{International Centre for Theory of Quantum Technologies, University of Gdańsk, Wita Stwosza 63, 80-308 Gdańsk, Poland}

\author{Bianka Woloncewicz}

\affiliation{International Centre for Theory of Quantum Technologies, University of Gdańsk, Wita Stwosza 63, 80-308 Gdańsk, Poland}

\author{Marek \.Zukowski}

\affiliation{International Centre for Theory of Quantum Technologies, University of Gdańsk, Wita Stwosza 63, 80-308 Gdańsk, Poland}

\begin{abstract}
We comment on the purported  violation of local realism, by a single photon induced correlations between homodyne detectors, which one can find in: 
New J. Phys. {\bf 10}, 113024 (2008), 	[arXiv:0807.0109]. The claim is erroneous, due to a calculational mistake. As the result is the basis of other claims in the paper, they are unsubstantiated.

\end{abstract}

\maketitle

\section{Introduction}


Ref. \cite{DUNNINGHAM} seems to be the first work to show that a single photon in a superposition of being in one or another exit of a 50-50 beamsplitter can induce correlations, in homodyne measurements at two spatially separated locations, which violate local realism. 
The work is an element of a long discussion concerning the suggested by Tan, Walls and Collett \cite{TWC91} ``non-locality of a single photon''. Ref.  \cite{DUNNINGHAM}  discusses essentially the same setup as \cite{TWC91}, Fig. 1. The difference is that in  \cite{TWC91} the local settings of the homodyne measurements were defined by the phases of local oscillator fields, whereas in \cite{DUNNINGHAM} the settings are defined by transmittivities of the beamsplitters at the final  local measurement stations. In both configurations, of \cite{TWC91} and \cite{DUNNINGHAM}, the (moduli of) amplitudes of the local oscillators are fixed, they do not change between settings, and are the same for Alice and Bob.  Ref. \cite{TWC91} also claimed a violation of local realism, however this was based on the authors' usage of Bell-like inequalities which involve an additional assumption, except the usual ones needed to derive Bell inequalities. Thus the claim of \cite{TWC91} is unfounded. A final proof of this is given in \cite{OurModel}, where one can find an explicit local realistic model for the original Tan-Walls-Collett correlations.

The attempt of Ref. \cite{DUNNINGHAM}  does not fall into this trap of incorrect Bell inequalities. They use the good old CHSH ones. Still the violation which they claim does not occur. There is an error in calculation. We show this below.

\subsection{Caveat}
 One may think that the question of ``nonlocality of a single photon'', for the configuration presented in \cite{TWC91}, was positively answered in the works of Banaszek and Wódkiewicz \cite{Banaszek99} and Hardy \cite{Hardy94}.
However both papers, to show violation of local realism, use measurement settings such that one of  them involves switched off  auxiliary field. Further, Ref. \cite{Banaszek99} uses displacement operations, instead of simple homodyning, and Ref.  \cite{Hardy94} gives a violation of local realism only for states which are a non-trivial superposition of single photon excitation, and the vacuum state. It does not work for just a single photon\footnote{In a forthcoming manuscript \cite{3rdPaper} we show that this is a robust feature: even for states with a vacuum component of norm of few percent, the Hardy method gives operationally inaccessible violations of a relevant Bell inequality. Simply,  they are very small.}. {\color{black}Thus the overall interferometric configurations \cite{Banaszek99} and \cite{Hardy94} differ very much with respect to \cite{DUNNINGHAM} and \cite{TWC91}, especially because in the case of the latter ones local oscillators have fixed power (it does not change with the change of measurement settings). In the case of 
\cite{Banaszek99} and \cite{Hardy94} one of the measurement settings is, with respect to the single photon state, a von Neumann observable (projector) and the other one is a POVM. In the case of \cite{DUNNINGHAM} and \cite{TWC91} for both settings we have POVM measurements.}

\subsection{Timeliness}
The question under what circumstances one can violate Bell inequalities with single photon induced correlations is  not an academic one anymore. With the current techniques realisations of the scheme of \cite{TWC91} are feasible, and as a matter of fact such experiments are being done, see e.g., the state-of-the-art-one \cite{WEAKHOMO-OPTICAL-COHERENCE}. The techniques of photon number  resolving homodyne measurement, as in \cite{WEAKHOMO-OPTICAL-COHERENCE}, allow experiments like the one of \cite{DUNNINGHAM}, as they {\em require} photon-number resolution.

\section{The setup}
\begin{figure}[t]
\includegraphics[width= 0.95 \columnwidth]{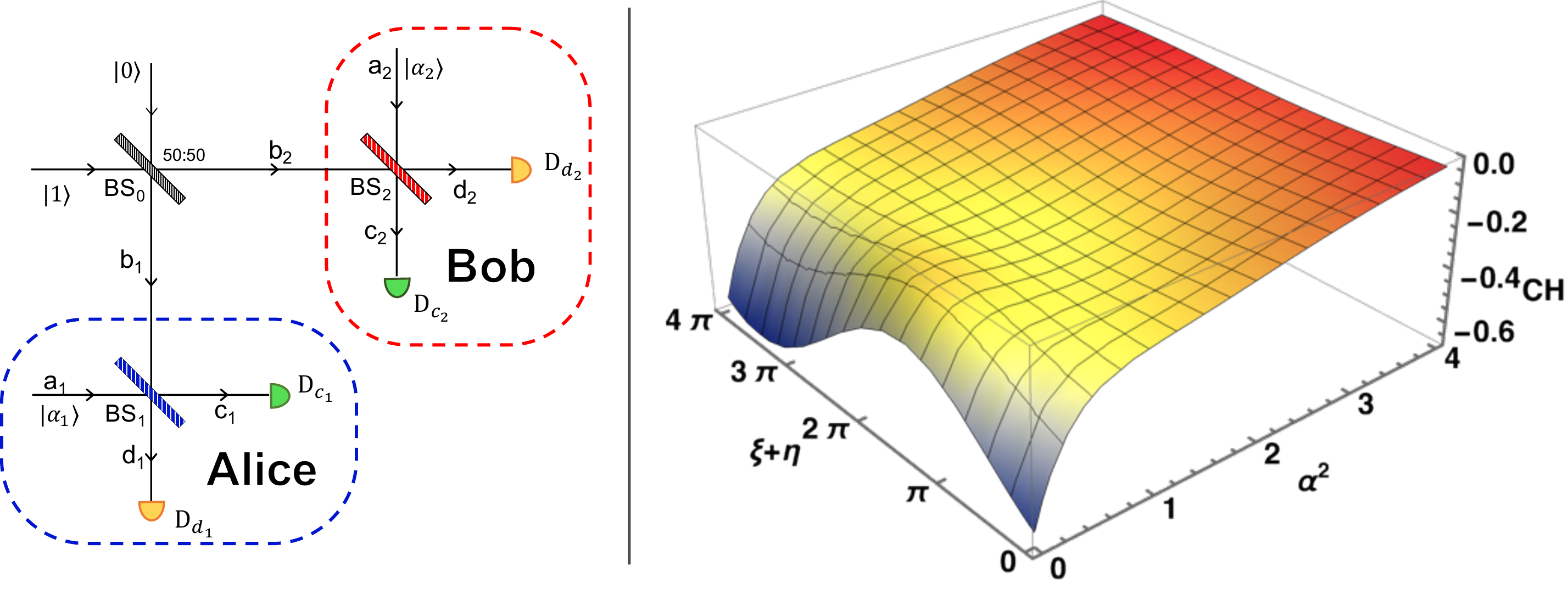}
	\caption{\label{mainSetup_CH}
		Left: Schematic diagram of the  experimental setup for uncovering the non-locality present in a single photon impinging on a balanced beamsplitter $BS_0$. The two output modes $b_1$ and $b_2$ then feed other two beamsplitters $BS_1$ and $BS_2$, of varying transmittivities, $\cos^2(\frac{\xi}{2})$ and $\cos^2(\frac{\eta}{2})$ respectively, which serve as local measurement settings. The other inputs of the variable beamsplitters are fed with local oscillators of fixed amplitude but different phases $\phi_1$ and $\phi_2$, with the constraint that the differences in phases $\Delta \phi = \phi_1 - \phi_2$ are fixed \cite{DUNNINGHAM}. Right: Plot of CH expression given in Eq. (\ref{CHfinal}), as a function of $\alpha^2$ and $\xi + \eta$, where the  other two parameters at $\Delta \phi = \frac{\pi}{2}$ and $\xi - \eta = \frac{3\pi}{4}$ are fixed to those values given in \cite{DUNNINGHAM}. There is no violation of CH inequality as the value lies between $(-1,0)$.}
\end{figure}

The experimental setup of Ref. \cite{DUNNINGHAM} is aimed at uncovering the nonlocality of a single photon that emerges from  a balanced beamsplitter $BS_0$ and it is shown in the left panel of the figure \ref{mainSetup_CH}. The output modes of $BS_0$ are connected with one of the input ports of other two beamsplitters $BS_1$ and $BS_2$ of variable transmittivity $\cos ^2\left(\frac{\xi }{2}\right) = T_1$ and $\cos ^2\left(\frac{\eta }{2}\right) = T_2$. Obviously, the reflectivities of the two beamsplitters are $\sin ^2\left(\frac{\xi }{2}\right)$ and  $\sin ^2\left(\frac{\eta }{2}\right)$.    The other two input ports $a_1$ and $a_2$ of $BS_1$ and $BS_2$ are fed with local oscillators $\ket{\alpha_1}_{a_1}$ and $\ket{\alpha_2}_{a_2}$, with $\alpha_1 = \alpha e^{i\phi_1}$ and $\alpha_2 = \alpha e^{i\phi_2}$, used to perform local homodyne measurements.
In the scheme of \cite{TWC91}   the transmittivities of the beamsplitters are fixed at $50\%$, that is we have $\xi=\eta=\frac{\pi}{2}$, and the detectors measure intensities. 
The authors of \cite{DUNNINGHAM} use as ``favorable'' detection  events single photon counts at detectors monitoring  $c_j$ accompanied with no counts at detectors in beams $d_j$, where $j=1,2$. 



The overall input state, which includes the local oscillators, before the two tunable beamsplitters, reads:
\begin{equation}
|\Psi\rangle=
\ket{\alpha e^{i \phi_1}}_{a_1}\frac{1}{\sqrt2}\left(|0,1\rangle_{b_1b_2}+i\,|1,0\rangle_{b_1b_2}\right)\ket{\alpha e^{i \phi_2}}_{a_2}, 
\label{PSIIN}
\end{equation}
and photon number resolving measurements are performed in the output modes $c_j$ and $d_j$. The favorable outcomes are ascribed value $-1$, and when any other event  occurs it gives  $+1$, \cite{DUNNINGHAM}.
The local measurement operators, for Alice and Bob, can therefore be written as: 
 \begin{eqnarray}
  \label{AliceM}
 A^\xi = \id - 2 (\ket{1,0}\bra{1,0})_{c_1d_1}, \quad
 B^\eta =\id - 2 (\ket{1,0}\bra{1,0})_{c_2d_2}, 
   \end{eqnarray}
 where $\xi$ and $\eta$ in the superscript denote the  parameters which tune the transmittivity of the variable beamsplitters $BS_1$ and $BS_2$, respectively, of Alice and Bob. 
Introducing the correlator:
\begin{eqnarray}
 E(\xi, \eta, \Delta \phi) = \sum_{i,j = \pm 1}ijP(i,j|\xi, \eta, \Delta \phi) = \bra{\Psi} A^\xi B^\eta \ket{\Psi},
\end{eqnarray}
where $P(i,j|\xi, \eta, \Delta \phi)$ is the joint probability of detecting events $i$ and $j$ for local settings $\xi$ and $\eta$, the authors of \cite{DUNNINGHAM} defined the CHSH inequality in the following form:
\begin{eqnarray}
\label{CHSHxi}
CHSH_{\xi\eta} = E(\xi, \eta, \Delta \phi) + E(\xi +\frac{\pi}{2}, \eta, \Delta \phi) - E(\xi, \eta +\frac{\pi}{2}, \Delta \phi) + E(\xi +\frac{\pi}{2}, \eta +\frac{\pi}{2}, \Delta \phi),
\end{eqnarray}
where the specific settings are those which are used and specified by \cite{DUNNINGHAM}. Note that $\Delta \phi$ is kept fixed, see \cite{DUNNINGHAM}.

\section{Analysis}
In the following analysis we utilise the equivalence between the CHSH inequality with  the Clauser-Horne (CH) inequality, which always holds provided we have only dichotomic results (observables),
which is clearly the case with the above value assignments.
This approach allows for clear identification of the error in the analysis of \cite{DUNNINGHAM}, which is slightly hidden when analysing solely the CHSH version.

The CHSH inequality in the general form reads
\begin{equation}
    \label{CHSHgen}
    CHSH=E(a,b)+E(a,b')+E(a',b)-E(a',b')\leq 2,
\end{equation}
and whenever the local measurements specified by settings $a,a'$ and $b,b'$ are dichotomic and have outcomes $\pm 1$ is equivalent to a CH inequality of the form \cite{ScaraniBook}:
\begin{equation}
    \label{CHgen}
    CH=P(-1,-1|a,b)+P(-1,-1|a',b)+P(-1,-1|a,b')-P(-1,-1|a',b')-P(-1|a)-P(-1|b)\leq 0.
\end{equation}
As can be seen the CH version depends solely on the probabilities of one chosen outcome under different settings. These will be the favorable outcomes, $-1$. The CHSH and the CH values are related by, see e.g. \cite{ScaraniBook}:
\begin{equation}
    \label{CHtoCHSH}
    CHSH=2+4CH.
\end{equation}

In ref. \cite{DUNNINGHAM}, the authors split the input state in Eq. (\ref{PSIIN}) as follows
\begin{equation}
|\Psi\rangle=
\Big(\alpha e^{-\alpha^2} \underbrace{\frac{1}{\sqrt{2}}(e^{i \phi_1} \ket{1,0}\ket{0,1} + i e^{i \phi_2} \ket{0,1}\ket{1,0} )}_{\ket{\psi_1}} + \sqrt{1 - \alpha^2 e^{-2\alpha^2}} \ket{\Lambda}\Big)_{a_1b_1a_2b_2}.
\label{PSIsplit}
\end{equation}
The state $\ket{\psi_1}$ is a maximally entangled two-qubit state. As such if considered alone it saturates the Tsirelson bound for optimally chosen measurements.



Let us focus on $\ket{\Lambda}$. 
Here, the authors of \cite{DUNNINGHAM},
erroneously claim that the CHSH value \eqref{CHSHxi} for the state $\ket{\Lambda}$ is always equal to $2$ regardless of the settings.
Their claim is equivalent to saying, that in $\ket{\Lambda}$ there is no term which contains single photon in part of Alice and at the same time more than one photon in part of Bob, or vice a versa. 
But this is actually not the case, 
since $\ket{\Lambda} $ e.g. contains  a term like $\ket{1,0}_{a_1b_1}\ket{1,1}_{a_2b_2}$, which leads to $-1$ in location of Alice and $+1$  in location of Bob. There are infinitely many  such terms. Hence the values of CHSH expression for  $\ket{\Lambda} $ is much less than $2$.  {\em In the following analysis we show explicitly, that the CHSH inequality evaluated on the entire state $\ket{\Psi}$ is never violated for the experimental scheme proposed by the authors of \cite{DUNNINGHAM}.}

The CH inequality for the event $-1$, which corresponds to the CHSH inequality \eqref{CHSHxi}, for the set of measurement settings  used in \cite{DUNNINGHAM} reads: 
\begin{eqnarray}\label{CH}
CH_{\xi\eta} = P(-1,-1|\xi, \eta, \Delta \phi) + P(-1,-1|\xi +\frac{\pi}{2}, \eta, \Delta \phi) - P(-1,-1|\xi, \eta +\frac{\pi}{2}, \Delta \phi) \nonumber \\
+  P(-1,-1|\xi +\frac{\pi}{2}, \eta +\frac{\pi}{2}, \Delta \phi)
- P(-1|\xi + \frac{\pi}{2},  \Delta \phi)  - P(-1| \eta, \Delta \phi).
\end{eqnarray}

For the input state $\ket{\Psi}$, given in Eq. (\ref{PSIIN}), the joint probability and the two local ones read respectively:  
\begin{eqnarray}
 &&P(-1,-1|\xi, \eta, \Delta \phi) = \frac{1}{4} \alpha ^2 e^{ -2 \alpha ^2} (1 - \cos (\eta ) \cos (\xi ) - \sin (\eta ) \sin (\xi ) \sin (\Delta\phi )), 
\\
&&P(-1|x,\Delta \phi) =  \frac{1}{2} e^{ -2 \alpha ^2} \left(\alpha ^2 \cos ^2\left(\frac{x }{2}\right)+\sin ^2\left(\frac{x }{2}\right)\right), ~~ x = \xi,\eta. 
\end{eqnarray}
Substituting these expressions in the CH expression (\ref{CH}), we get:  
\begin{eqnarray}
    CH_{\xi\eta}(\ket{\Psi}) &=& \frac{1}{4} e^{-2 \alpha ^2} \Big(\alpha ^2 ( 1+ \sin \Delta \phi ) (\sin ( \xi - \eta) - \cos ( \xi - \eta)) \nonumber \\
    && +~ e^{\alpha ^2} \left(1 - \alpha ^2\right) (\cos \eta -\sin \xi )+2 \alpha ^2-2 e^{\alpha ^2} \left(\alpha ^2+1\right)\Big), \label{CHfinal}
\end{eqnarray}
in which $CH_{\xi\eta}(\ket{\Psi})$ indicates that the expression is evaluated for the state $\ket{\Psi}$.
Note that, as we have \eqref{CHtoCHSH} the CHSH value also depends on both $\xi - \eta$ and $\xi + \eta$, in contrast to what is stated  in  Eq. (13) of \cite{DUNNINGHAM}. 

Let us choose the values of the parameters used in \cite{DUNNINGHAM}: $\Delta \phi = \frac{\pi}{2}$ and $\xi - \eta = \frac{3\pi}{4}$. 
We plot the CH value (\ref{CHfinal}) as a function of $\alpha^2$ and $\xi + \eta$ in the right panel in figure \ref{mainSetup_CH}. As can be seen there is no violation of the CH inequality for any values of $\alpha^2$ and $\xi + \eta$, and therefore the corresponding CHSH inequality \eqref{CHSHxi} is also not violated in contradiction with the claim of \cite{DUNNINGHAM}. The 
CHSH expression evaluated for $\ket{\Psi}$ reads: 
\begin{eqnarray}\label{CHSH}
    CHSH_{\xi\eta}(\ket{\Psi})   &=& 2 +  e^{-2 \alpha ^2} \Big(\alpha ^2 ( 1+ \sin \Delta \phi ) (\sin ( \xi - \eta) - \cos ( \xi - \eta))  \nonumber \\ 
    &&  +~ e^{\alpha ^2} \left(1 - \alpha ^2\right) \left(\cos \eta -\sin \xi \right)+2 \alpha ^2-2 e^{\alpha ^2} \left(\alpha ^2+1\right)  \Big). 
\end{eqnarray}
It {\em never goes beyond the value of $2$}, even for an overall optimization of all the parameters.

\mm{It is worth to emphasise that the analysis of the Bell setup with dichotomic outcomes is much easier when conducted with the use of CH instead of CHSH inequality, since then one needs to focus only on one chosen (here, favorable) local  outcome. In the CHSH analysis one needs to analyse correlations between all pairs of outcomes, which, as more cumbersome,  can lead to  mistakes.}

\section{Discussion}
In the analysis of \cite{DUNNINGHAM}, the authors assumed fixed coherent local oscillator strength for both local measurement settings of Alice and  Bob. Moreover, the phases of the coherent states are also the same for both settings and the difference between the phases of the local oscillators of Alice and Bob has been fixed to $\frac{\pi}{2}$. 
They also use local setting of $\xi$ and $\eta$ which differ locally by fixed $\pi/2$.
We have relaxed some of these constraints concerning the measurement settings \cite{DUNNINGHAM}. The results are :
\begin{itemize}
    \item If the transmittivity of the beamsplitters, ($\xi$ and $\eta$), and the phases of the coherent states ($\phi_1$ and $\phi_2$)  vary locally and may be different among the parties, and the strengths of the coherent states $\alpha_1$ and $\alpha_2$ are different for Alice and Bob but the same for the two local settings, we observed no violation.
\end{itemize}
\mm{The above analysis strongly suggests that in order to obtain a Bell violation in the considered experimental setup one should locally vary the local oscillator strengths of the beamsplitters from setting to setting. This is fully concurrent with our analysis of the setup presented in \cite{OurModel, 2ndPaper} and in the forthcoming work \cite{3rdPaper}. 

 As a last remark we note that the authors of \cite{DUNNINGHAM} draw a lot of conclusions from the purported violation of the Bell inequality, concerning the possibility of performing the experiment with entirely independent local oscillators with uncontrollable phases. These conclusions are not substantiated since the basis of their reasoning is entirely wrong --- there is no violation of the CHSH inequality for the proposed measurement setup. We leave as an open question whether their conclusions hold also in the case of  situations which correctly lead to violations of local realism.}
 
\textbf{Acknowledgments}
Work supported by  Foundation for Polish Science (FNP), IRAP project ICTQT, contract no. 2018/MAB/5, co-financed by EU  Smart Growth Operational Programme. MK is supported by  FNP  START scholarship. AM is supported by (Polish) National Science Center (NCN): MINIATURA  DEC-2020/04/X/ST2/01794.
   
\bibliography{SinglePhoton}

\end{document}